\documentstyle[amsfonts,epsf]{europhys}

\def\stars{\bigskip\centerline{***}\medskip}

\newif\ifboo \boofalse

\def\Review#1{\boofalse{\it #1},}
\def\Name#1{{\sc #1},}
\def\Vol#1{\ifboo Vol. {\bf #1}\else{\bf #1}\fi}
\def\Year#1{\ifboo #1\else(#1)\fi}
\def\Book#1{\bootrue{\it #1},}
\def\Page#1{\ifboo {\rm p. #1}\else{\rm #1}\fi}

\newcommand{\nc}{\newcommand}
\newcommand{\cft}{conformal field theory }
\newcommand{\cor}{\overline{\langle \sigma(0)\sigma(R)\rangle^p}}
\newcommand{\opx} {{\cal O}_p (x)}
\newcommand{\nn}{\nonumber}

\nc{\beq}{\begin{equation}}
\nc{\eeq}{\end{equation}}
\nc{\beqa}{\begin{eqnarray}}
\nc{\eeqa}{\end{eqnarray}}
\nc{\eps}{\epsilon}
\nc{\s}{\sigma}
\nc{\veps}{\varepsilon}
\nc{\no}{\noindent}
\nc{\D}{\Delta}
\nc{\al}{\alpha}
\nc{\be}{\beta}
\nc{\ga}{\gamma}
\nc{\de}{\delta}
\nc{\dmi}{{1\over 2}}

\begin{document}
%
%
%
\euro{}{}{}{}
\Date{}
\shorttitle{M.-A LEWIS Further evidence of the absence of RSB in random
bond Potts models}
%
%
%
\title{Further evidence of the absence of Replica Symmetry Breaking in Random Bond
Potts Models}
\author{Marc-Andr\'e Lewis\footnote{E-mail:lewism@lpthe.jussieu.fr}}
\institute{Laboratoire de Physique Th\'eorique et des Hautes Energies, \\
         Universit\'es Pierre et Marie Curie (Paris VI) et Denis Diderot (Paris VII),\\
Boite 126, Tour 16, 1er \'etage,\\         
4 pl. Jussieu, 75251 Paris CEDEX 05, FRANCE}
     
%
%
\rec{}{}
%
%
%
\pacs{
\Pacs{64}{60Ak}{Renormalization group studies of phase transitions}
      }
\maketitle

%
%
%
\begin{abstract}
In this short note, we present supporting evidence for the replica symmetric
approach to the random bond $q$-state Potts models.  The evidence is
statistically strong enough to reject the applicability of the Parisi
replica symmetry breaking scheme to this class of models.  
The test we use is a generalization of one formerly proposed by Dotsenko et
al. \cite{DDP} and consists in measuring scaling laws of disordered-averaged
moments of the spin-spin correlation functions.  Numerical results, obtained
via Monte Carlo simulations for several values of $q$, are shown to be in
fair agreement with the replica symmetric values computed by using perturbative CFT
\cite{DDP,MAL} for the second and third moments of the $q=3$ model.   RSB
effects, which should increase in strength with moment, are unobserved. 
\end{abstract}
%
%
%
Since its first application to the study of glassy systems, the replica
approach has been a useful tool in the disordered models research field.
However, as was rapidly observed, its straightforward application (that is,
assuming all replicas are identical) to physical systems can lead to
serious problems, the most notorious being the negative entropy it gives for
the spin glass model. These non-physical conclusions can be avoided by
breaking the replica symmetry.  The way in which the symmetry has to be
broken in the spin glass problem was first explained by Parisi \cite{MPV}.
However, there are systems in which the symmetry is not broken and where
the replica symmetric (RS) approach is valid.  There is no straightforward
arguments to decide whether a system studied using the replica method
exhibits replica symmetry breaking (RSB) or not.  

The aim of this short note is to implement a test that can reveal the
presence of RSB in disordered local-interaction spin
systems.  We shall consider disordered Potts models, where disorder is
introduced via randomness in bond strengths.  This problem was
studied perturbatively using the replica technique \cite{LU,DPP}.  There,
critical exponents were computed assuming RS and the
obtained values were shown to be in agreement with Monte Carlo data.
However, a possible RSB would not affect those
quantities significantly and the apparent agreement couldn't rule out its
existence.

Recently, Dotsenko et al. \cite{DDP} proposed a test of RSB
for random bond Potts model.  There, they showed that eventual RSB
effects could be observed if one considered the disorder-averaged moments
of spin-spin correlation functions.  They studied the second moment and
agreement was found with RS solutions, thus rejecting
RSB-induced deviations that would have been greater than the obtained
statistical resolution.  

In a recent letter, we exposed perturbative CFT computations of the $p$-th
moment in the replica symmetric case.  This naturally proposed the search
for signs of RSB in these correlators.  Although we have not computed
explicitely the deviations (this would involve a tensorial formulation of
the Parisi scheme), one can easily convince himself that they should become
more and more important as $p$ increases.  Performing Monte-Carlo
simulations, we computed the third moment for $3,4$ and $8$-state Potts
Model. For all these models, we observe scaling laws, thus showing that
there is no RSB.  The associated critical exponents are shown to be in fair
agreement, for the 3-states model, with CFT predictions, although the
perturbative expansion is not expected to be very precise for $q=3$.  This is the first
validation of our formula, previously exposed \cite{MAL}, which goes a order
further in perturbation than the one originally given by Ludwig \cite{LU}.

\section{Perturbative CFT results}
We shall not repeat here the renormalization group computations of higher
moments, which can be found, although not in details, in references
\cite{DDP,MAL}.  We rather give a short overview, only stating relevant
results. 

The partition function of the nearly-critical $q$-states random bond Potts
model, is well known to be of the form
\begin{equation} 
\label{Z}
Z(\beta) =  \mbox{Tr } \exp\{-H_0-H_1\},
\end{equation}
where $H_0$ is the Hamiltonian of the \cft corresponding to the $q$-states
Potts model with coupling constant $J_0$ the same for each bond.  The
Hamiltonian $H_1$, being the deviation from the critical point induced by
disorder is of the form
\begin{equation}
H_1= \int d^2x \,\tau(x)\epsilon(x),
\end{equation}
where $\tau(x)\sim\beta J(x) - \beta_c J_0$ is the random temperature
parameter.  The theory is defined on the whole plane.  We shall assume, for
simplicity, that $\tau(x)$ has a gaussian distribution for each $x$, with
\begin{eqnarray} 
\overline{\tau(x)} &=& \tau_0 \,=\frac{\beta-\beta_c}{\beta_c}\\
\overline{(\tau(x)-\tau_0)(\tau(x')-\tau_0)} &=& g_0\, \delta^{(2)}(x-x').
\end{eqnarray}

The usual way of averaging over disorder is to introduce replicas, that is, $n$ identical copies of the same model, for which:
\begin{equation}
(Z(\beta))^n=\mbox{Tr }\exp\left\{-\sum_{a=1}^n H_0^{(a)} - \int d^2x
\,\tau(x)\sum_{a=1}^n \varepsilon_a (x)\right\}.
\end{equation}
Taking the average over disorder by performing gaussian integration, one gets
\begin{equation}
\overline{(Z(\beta))^{n}}=\mbox{Tr
}\exp\left\{-\sum_{a=1}^{n}H_{0}^{(a)}- \tau_{0}\int
d^{2}x\sum_{a=1}^{n}\varepsilon_{a}(x) + g_{0}\int
d^{2}x\sum_{a\neq b}^{n}\varepsilon_{a}(x)\varepsilon_{b}(x)\right\}.
\end{equation}
This is a field theory of $n$ coupled models with coupling action given by
\begin{equation}
H_{\mbox{int}}=-g_{0}\int d^{2}x\sum_{a\neq
b}^{n}\varepsilon_{a}(x)\varepsilon_{b}(x). 
\end{equation}
Only non-diagonal terms are kept since diagonal ones can be
included in the Hamiltonian $H_0$.  Moreover, they can be shown to have
irrelevant contributions, since their OPE consist of the identity plus
terms that are irrelevant at the pure fixed point. We now turn our
attention to the $p$-th moment of the spin-spin correlation function
$\cor$.  In terms of replicas, it can be written as
\begin{equation} 
\cor = \lim_{n\rightarrow 0}
\frac{(n-p)!}{n!\,p!}  \left\langle \sum_{a_1\ne\cdots\ne a_p}^n
\sigma_{a_1}(0)\cdots \sigma_{a_p}(0) \sum_{b_1\ne \cdots \ne b_p}^n
\sigma_{b_1}(R)\cdots\sigma_{b_p}(R)\right\rangle
\end{equation} 
Thus, the operator to be renormalized is 
\begin{eqnarray} 
{\cal O}_p(x)&\equiv&\sigma_{a_1}(x)\sigma_{a_2}(x)\cdots\sigma_{a_p}(x)\\
&&\,\,a_1 \ne a_2\cdots \ne a_p,\,\, 1\le a_i \le n,
\end{eqnarray} 
perturbed by the interaction term
\begin{eqnarray} 
\tilde\opx &\equiv& {\cal O}_p \exp\{-H_{\mbox{int}}\} \\
&=& {\cal
O}_p \left( 1-H_{\mbox{int}}+\frac{1}{2}(H_{\mbox{int}})^2-\cdots\right).
\end{eqnarray} 
Renormalization group computations lead to the identification of a
non-trivial fixed point, at which we are able to compute the correlation
functions.  Using scaling laws, we get 
\begin{eqnarray} 
\cor &\sim& \lim_{n\rightarrow 0}
\frac{(n-p)!}{n!}\sum_{a_1\ne a_2 \cdots \ne a_p} (Z(\xi_R))^2
\frac{1}{R^{2p\Delta_{\sigma}}}\nn \\ &\sim&
\frac{(Z(\xi_R))^2}{R^{2p\Delta_{\sigma}}}. 
\end{eqnarray} 

The final result is obtained by using the fixed point value $Z(\xi_R) \sim
e^{\gamma_* \xi_R} = R^{\gamma_*}$.  The RG study introduces a parameter
$\epsilon$, which can be seen as proportional to the central charge
deviation of the pure model from the Ising value of $1/2$.  For the
$3$-state Potts model, $\epsilon=2/15$.  For generic $\epsilon$, one gets
(in \cite{MAL}, $\alpha$ should be replaced by $-\alpha$):
\begin{equation} 
\cor \sim \frac{1}{R^{2p\Delta_{\sigma^p}'}}. 
\end{equation} 
with
\begin{equation} 
\Delta_{\sigma^p}' = \Delta_{\sigma} - \gamma_*(p),
\end{equation}
\begin{equation}
\gamma_*(p) = \frac{9}{32}(p-1)\left(\frac{2}{3}\epsilon +
\left(\frac{11}{12}-\frac{2 K}{3}+\frac{\alpha}{24}(p-2)\right)
\epsilon^2\right)+ {\cal O}(\epsilon^3),
\end{equation}
and
\begin{equation}
K=6\log 2 \qquad \alpha = 33-\frac{29\sqrt{3}\pi}{3}.
\end{equation}

Thus, perturbed conformal field theory predicts, for the $3$-state Potts
models, the following values for the second and third moments:
\begin{equation}
2\Delta_{\sigma^2}' = \frac{4}{15} - 0.0314 = 0.235
\end{equation}
\begin{equation}
2\Delta_{\sigma^3}' = \frac{4}{15} - 0.0466 = 0.220
\end{equation}

\section{Monte Carlo Simulations}
To search for signs of RSB, and, in the absence of it, to confirm RS
values, we performed Monte Carlo simulations of the random bond $q$-Potts
model for $q=2,4,8$.  The method used follows the one in \cite{DDP}.
To study scaling effects on the correlation functions, we studied square
lattices of side $L$ ranging from $10$ to $500$.  Since we wanted to
exhibit a possible break of the replica symmetry, the algorithm has to be
chosen in such a way that it doesn't assume the symmetry a priori.  For
this reason, we simulated three configurations of the $q$-Potts model with
same disorder, but different initial conditions and independent
thermalizations.  We computed the products of magnetization
\begin{equation}
Q_3= \frac{1}{L^2} \sum_{i=1,L^2} \langle \sigma_i^a\rangle \langle
\sigma_i^b\rangle\langle\sigma_i^c\rangle,
\end{equation}    
and
\begin{equation}
Q_2= \frac{1}{L^2} \sum_{i=1,L^2} \langle \sigma_i^a\rangle \langle
\sigma_i^b\rangle,
\end{equation}
with $\langle\sigma_i^a\rangle$ being the thermal average of the local
magnetization 
\begin{equation}
\sigma_i^a \equiv \vec\sigma_i^a \cdot \vec m^a,
\end{equation}
where $\vec m^a$ is the mean magnetization of lattice $a$;
\begin{equation}
\vec m^a = \frac{1}{L^2} \sum_{i=1,L^2} \vec\sigma_i^a.
\end{equation}
It is rather obvious, since all lattices are thermalized independently, that
$Q_{3}$ and $Q_2$ are indeed the same as
\begin{equation}
Q_3= \frac{1}{L^2} \sum_{i=1,L^2} \langle \sigma_i^a\sigma_i^b\sigma_i^c\rangle
\end{equation}
\begin{equation}
Q_2= \frac{1}{L^2} \sum_{i=1,L^2} \langle \sigma_i^a\sigma_i^b\rangle.
\end{equation}

Measurement were performed on square lattices with toroidal boundary
conditions.  The Hamiltonian of the simulated model is
\begin{equation}
H=-\sum_{\{i,j\}}J_{ij}
\left(\delta_{\sigma_i^a,\sigma_j^a}+\delta_{\sigma_i^b,\sigma_j^b}+
\delta_{\sigma_i^c,\sigma_j^c}\right),
\end{equation}
where we took the coupling between nearest neighbours to be
\begin{equation}
J_{ij} = J_0 \mbox{   or   } J_1 
\end{equation}
with equal probabilities.  This makes it possible to make the system
self-dual by tuning the temperature so that the relation
\begin{equation}
\frac{1-e^{-\beta J_0}}{1+(q-1)e^{-\beta J_0}} = e^{-\beta J_1}
\end{equation}
is obeyed.  We chose $J_0/J_1 = 10$ for the simulations with $q=3,4$, which
is strong enough to avoid cross-over effects \cite{Picco,Cardy}.  For the $q=8$
model,we rather chose $J_0/J_1 = 8.5$, again because this seems the
appropriate value to avoid cross-over and minimize the spread of our data set.

Autocorrelation times were coarsely evaluated and the statistics ajusted in
such a way that thermal fluctuations can be ignored (typically,
for a single disorder configuration, thermalization period was at least 70
auto-correlation times long and at least 200 measures were taken (one every
auto-correlation time).  To average over disorder, we made measurements for
20 000 disorder configurations (10000 for $q=8$).  Doing so, one can
extract critical exponents straightforwardly:
\begin{equation}
\overline{Q_p} = K L^{-p\Delta_{\sigma^p}'},
\end{equation}
where $K$ is a non-universal constant.  The exponent can then be obtained
by taking logarithms.
     
The results of our simulations, shown in Figures 1, 2 and 3, clearly
support the RS scenario.  In these figures, we present log-log plots of
$\overline{Q_p}^{2/p}$ versus $L$ ($p=2,3$), for the three, four and
eight-state Potts models.  By taking the slopes of these graphs, we can
extract $2\Delta_{\sigma^p}'(q)$ (which is minus the slope).  None of the models presented show
significant deviations from scaling which should arise if the replica
symmetry was broken.

For the 3-state Potts model, the critical exponents associated to the
scaling behaviour are in fair agreement with the values predicted by
perturbative CFT computations.  The deviations from the pure model behaviour are:
\begin{eqnarray}
2\gamma^*(2) &=&   0.0387 \qquad \mbox{Monte Carlo} \\
           &=&     0.0314  \qquad \mbox{CFT prediction}
\end{eqnarray}
\begin{eqnarray}
2\gamma^*(3) &=&   0.0648 \qquad \mbox{Monte Carlo} \\
             &=&   0.0466 \qquad \mbox{CFT prediction}
\end{eqnarray}
The numerical agreement is indeed quite surprising, especially for the third
moment, where the perturbative expansion is near the end of its
validity region \cite{MAL}.  Olson and Young \cite{OY} also computed
spin-spin correlation functions moments, but in a different optic and with
a different method.  Our values for the exponents, presented in Table I,
are in fair agreement with theirs, altough they seem to be systematically
lower.  Using other methods, Palagyi, Chatelain et {\em al.} \cite{CB} obtain values
that confirm this discrepancy.  Our values are equal to theirs, within
statistical errors.    
\begin{center}
\begin{figure}
\begin{center}
\leavevmode
\epsfxsize=3in\epsfbox{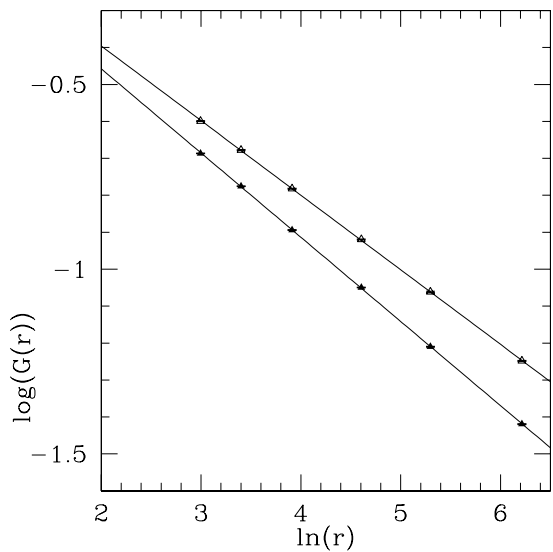}
\end{center}
\caption{Log-Log plot of $\overline{Q_p}^{2/p}$ with $p=2$ (lower line)
and $p=3$ (higher line)
for the random $3$-state Potts model}
\label{3state}
\end{figure}
\begin{figure}
\epsfxsize=3in
\begin{center}
\leavevmode
\epsfbox{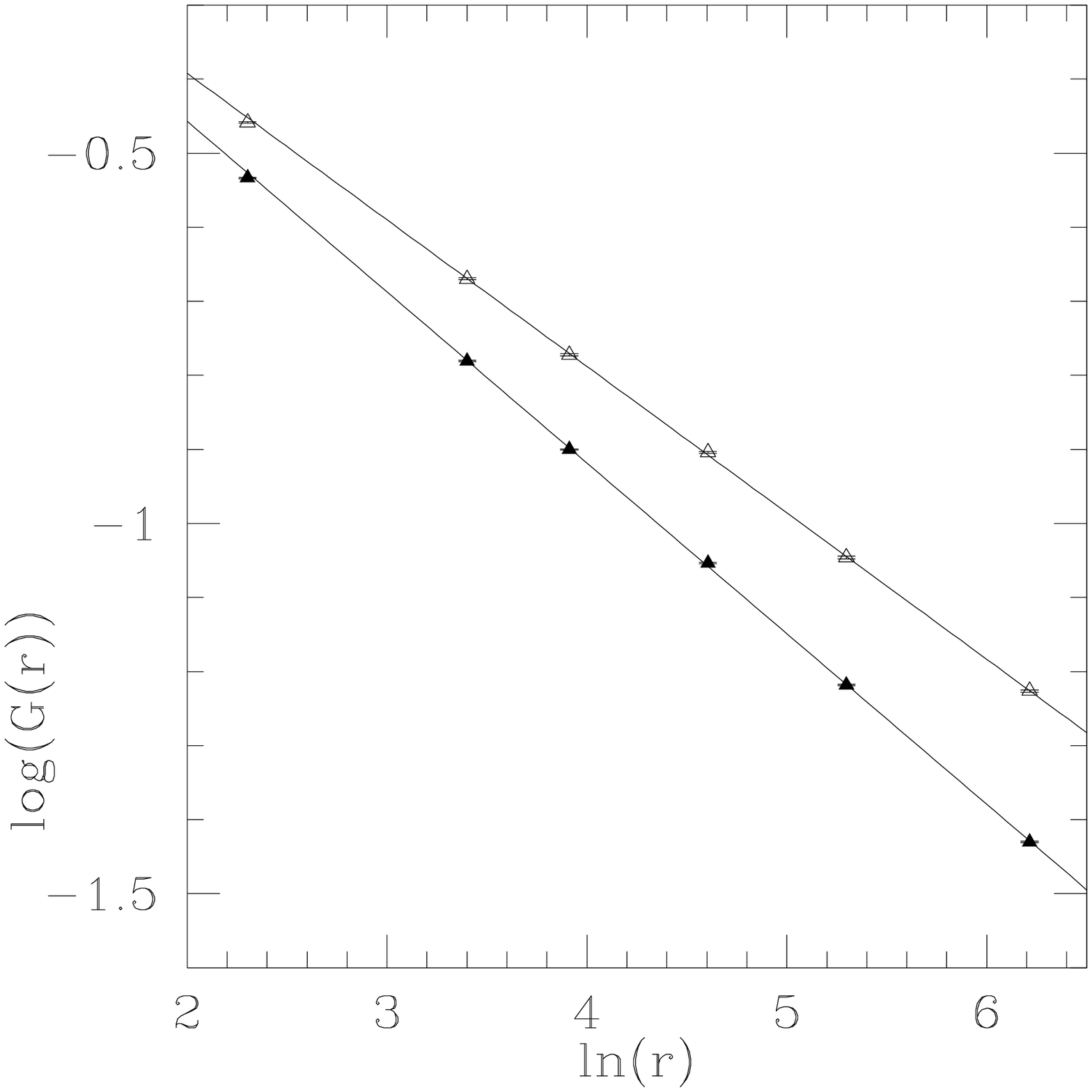}
\end{center}
\caption{Log-Log plot of $\overline{Q_p}^{2/p}$ with $p=2$ (lower line)
and $p=3$ (higher line)
for the random $4$-state Potts model}
\label{4state}
\end{figure}
\begin{figure}
\epsfxsize=3in
\begin{center}
\leavevmode
\epsfbox{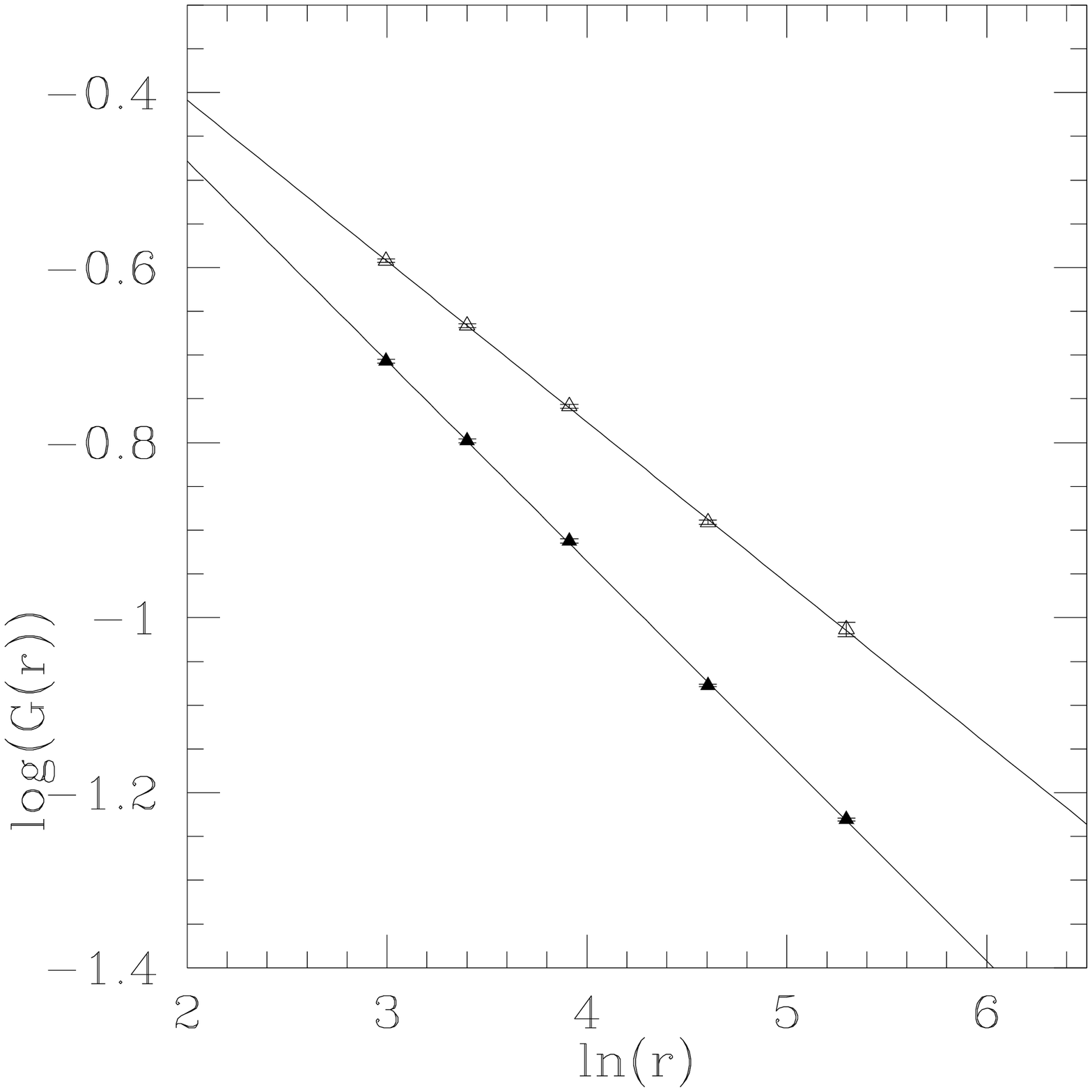}
\end{center}
\caption{Log-Log plot of $\overline{Q_p}^{2/p}$ with $p=2$ (lower line)
and $p=3$ (higher line)
for the random $8$-state Potts model}
\label{8state}
\end{figure}
\end{center}

\begin{table}
\begin{center}
\begin{tabular}{|c|l|l|} \hline
$\,\,\,q\,\,\,$ & $2\Delta_{\sigma^2}'$ & $2\Delta_{\sigma^3}'$ \\ \hline
$\,\,\,3\,\,\,$   &  $0.228(1)$	 & $0.202(2)$ \\ \hline	
$\,\,\,4\,\,\,$   &  $0.231(2)$  & $0.198(2)$ \\ \hline
$\,\,\,8\,\,\,$   &  $0.229(2)$ & $0.184(3)$ \\ \hline
\end{tabular}
\end{center}
\caption
{The values of the spin exponents $2\Delta_{\sigma^2}'$ and $2\Delta_{\sigma^3}'$, for
$q=3,4,8$. The normalisation was chosen so that exponents for the same value
of $q$ could be compared. The number in parentheses is the statistical
error in the last decimal place}
\end{table}	
\section{Conclusion}
We believe that the presented evidence is enough to rule out the RSB
scenario in random bond Potts models. If this symmetry was broken following
Parisi's scheme, deviations from the observed scaling laws would be, for the second moment, of the order of
10\%, and thus would be easily observed.  One can convince himself that the
deviation should become more apparent for the third moment, something which
is clearly not observed.

It will be interesting to see how more precise numerical methods, such as
transfer matrices iterations \cite{JC}, could give accurate values for
moments via cumulant expansions (for integer and non-integer values of $q$).  

\stars
We would like to thank Vl.S. Dotsenko, M. Picco, J.L. Jacobsen,
C. Chatelain and B. Berche for helpful comments and suggestions.  We also
acknowledge financial support from the NSERC Canada Scholarship Program.

\end{document}